# Analysing high resolution digital Mars images using machine learning

**Gergácz Mira** [1,2,3,4]  **Kereszturi Ákos** [3,5]

[1] *Luleå University of Technology, Bengt Hultqvists väg 1, 981 92 Kiruna, Sweden*
[2] *Wigner RCP, H-1121 Budapest, Konkoly-Thege Miklós 29-33, Hungary*
[3] *Konkoly Thege Miklos Astronomical Institute, Research Centre for Astronomy and Earth Sciences, H-1121 Budapest, Konkoly-Thege Miklós 15-16, Hungary*
[4] *ELTE Institute of Physics, H-1117 Budapest, Pázmány Péter 1/A, Hungary*
[5] *CSFK, MTA Centre of Excellence, H-1121 Budapest, Konkoly-Thege Miklós 15-17, Hungary*

## Abstract

The search for ephemeral liquid water on Mars is an ongoing activity. After the recession of the seasonal polar ice cap on Mars, small water ice patches may be left behind at shady locations due to the low thermal conductivity of the Martian surface and atmosphere. During late spring and early summer, these patches may be exposed to direct sunlight and warm up rapidly enough for the liquid phase to emerge.

To see the spatial and temporal occurrence of such ice patches, large number of optical images should be searched for and checked. Previously a manual image analysis was conducted on 110 images from the southern hemisphere, captured by the High Resolution Imaging Science Experiment (HiRISE) camera onboard the Mars Reconnaissance Orbiter space mission. Out of these, 37 images were identified with smaller ice patches, which were distinguishable by their brightness, colour and strong connection to local topographic shading.

In this study, a convolutional neural network (CNN) is applied to find further images with potential water ice patches in the latitude band between -40° and -60°, where the seasonal retreat of the polar ice cap happens. The previously analysed HiRISE images were used to train the model, where each image was split into hundreds of pieces (called chunks), expanding the training dataset to 6240 images. A test run conducted on 38 new HiRISE images indicates that the program can generally recognise small bright patches, however further training might be needed for more precise identification. This further training has been conducted now, incorporating the results of the previous test run. To retrain the model, 18646 chunks were analysed and 48 additional epochs were ran.

Using a CNN model may make it realistic to analyse all available surface images, aiding us in selecting areas for further investigation. After retraining the model, it produced a 94% accuracy in recognising ice patches in HiRISE images, 58% of these images showed small enough ice patches on them, while the rest of the images was covered by too much ice or showed $CO_2$ ice sublimation in some places.

# 1 Introduction

The aim of this work is to test the automated detection of surface ice patches on images of the Martian surface, using a convolutional neural network (CNN) by teaching the system using manually selected proper images.

Understanding the possibilities of ephemeral liquid water occurring on the surface of Mars under present day conditions is a main interest in Mars research. As the seasonal polar ice caps recede, small ice patches may remain in shaded areas after the sublimation of CO2 in the surroundings. During summer as they met with direct sunlight, irradiance increases and if the temperature is to rise fast enough, the possibility of the liquid phase appearing for a short time occurs [1]. The appearance of such ice patches was studied in this work, using high resolution images taken by the High Resolution Imaging Science Experiment (HiRISE) camera on board of the Mars Reconnaissance Orbiter (MRO).

Due to the low thermal conductivity and inertia of the Martian surface and atmosphere [2, 3], it is possible that small ice patches [4] may remain on the surface subsequent after the retreat of the seasonal polar cap in locations where they are shielded from the direct sunlight, for example by slope angles and exposure directions e.g. on the poleward side of elevated shadowing surface features. Over time water ice in these protected areas may also be exposed to direct sunlight, and then the ice may warm rapidly – it is not yet known whether a liquid phase [4, 5] may then appear, which is an important question for chemical transformations and the potential for life [6, 7, 8].

If the liquid phase emerges, it might influence slow, low temperature chemical changes on Mars, especially if supported by subzero temperature microscopic liquid water like proposed for hydrogen-peroxide decomposition [9] or for sulphate formation [10]. Such locations might need focused analysis in the future by orbiters monitoring them, which requires specific information on their location, the time period in which ice is present there and a selection of the best ones among them regarding potential chemical changes.

The selection of images with ice patches requires much manual work, as they should be separated from bright rocks, slopes with solar facing exposure direction, and large part of the surface might be ice of bright dust covered, beside the existence of other unique surface features, occasionally recorded under non proper conditions, and even influenced by hazes or clouds.

# 2 Methods

During the work manually selected images were used to teach the system what type of surface features should be searched for by a convolutional neural network.

## 2.1 The HiRISE camera

The MRO spacecraft has been orbiting Mars since 2006 with the HiRISE camera on board and had optically covered about 4% of the surface as of 2021, with many locations imaged repeatedly. The high-resolution camera has a mirror telescope with a diameter of 0.5 metres, making it the largest one ever used around another planet. From an altitude of 300 kilometres, it can achieve a pixel size of 25 centimetres, making it possible to survey the surface in great detail. The overall image size is 6 kilometres (20 000 pixels) in width by a programmable image length of up to 60 kilometres (200 000 pixels). It captures images between 14-16 local time and produces colour images in the central portion of the field of view in three wavelength bands: 400-600 nm (blue-green, B-G), 550-850 nm (red, R) and 800-1000 nm (near infrared, NIR).

### 2.2 Surveyed region and dataset

Previously a manual survey has been conducted on HiRISE images using the JMars software [11, 12]. The area of interest was the latitude band between -40° and -70° in the seasonal range of 140°-200° solar longitude, when the seasonal polar ice cap is receding in the southern hemisphere. These images were analysed by eye and the surface features were characterised based on scientific publications and personal knowledge gained during the work. These images were then categorised into groups with and without sufficient remnant ice patches. Such features were categorized as an ice patch, which were lighter than the surrounding surface, didn't cast shadows and were located in shaded areas of elevated surface features.

As of 2022, out of the approximately 1400 available HiRISE images that met the selection criteria, 110 ones were analysed manually. Potential residual ice patches were identified on 37 images from this dataset, which were used to teach the system for identification of similar patches on a larger number of HiRISE images.

### 2.3 Convolutional Neural Network (CNN)

So far the image analysis has been done manually by eye, which is quite a time consuming task given the size and number of the images. Therefore, increasing the dataset and potentially analysing all of the available images would be achievable using an automatized system, which is being attempted in this study.

CNNs are successful in image classification tasks, while also being able to efficiently process large datasets. They extract local patterns and features by applying convolutional layers. These layers consist of convolutional kernels which slide over the input scanning the image, and by convolving these with the input image, textures, edges and gradients are extracted. The convolutional layers produce a feature map, which represents the activation of each kernel at different spatial locations. By applying a nonlinear activation function to that, the learning process becomes nonlinear and the network becomes able to recognise more complex connections between the detected features.

After convolution, pooling operations were applied to reduce the spatial dimensions of the feature maps and summarise the information in a local area, thus improving computational efficiency and enabling the network to focus on the presence of a feature, rather than the exact position of it in the input image.

## 3   The neural network

In this work a smaller Xception network [13] was used to tackle the problem. This type of network, unlike traditional CNNs that use convolutional layers sequentially, it uses depth wise separable convolution. This means that the convolution operation is separated into two main steps, thus reducing the operations required and increasing the computational efficiency. The structure of the applied model can be seen in Figure 1.

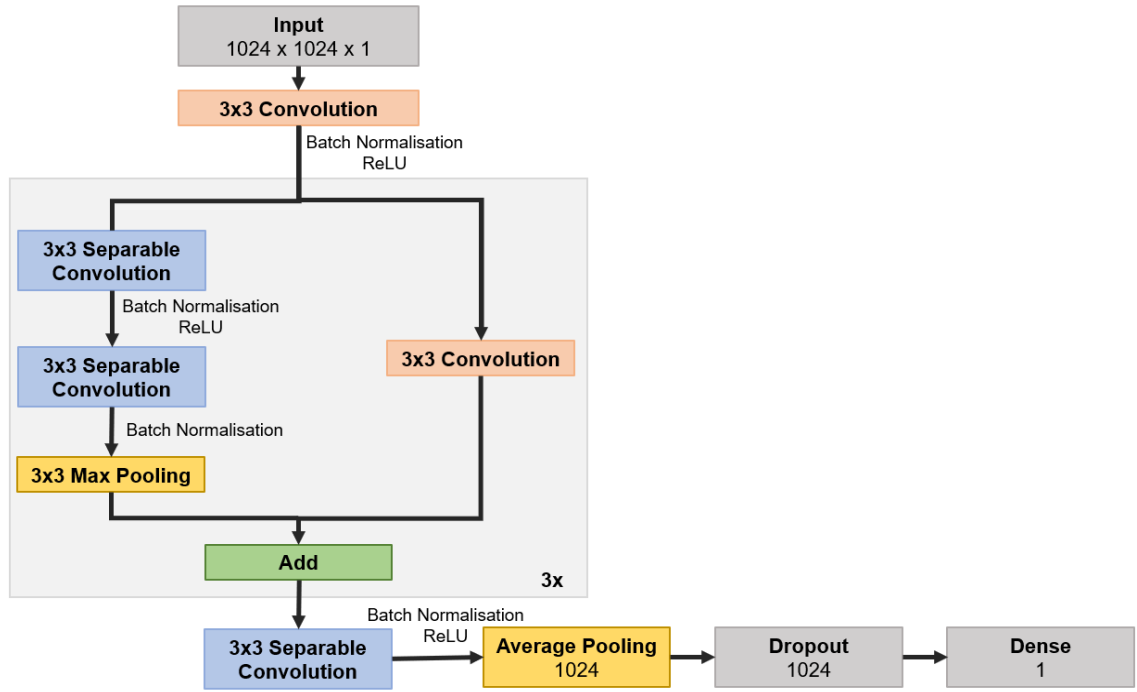

**Figure 1.** The structure of the model. Before each ReLU activation a Batch Normalisation is applied, to normalise the inputs to the activation and thus improve learning.

The model was running on a batch size of 25 through 25 training epochs at first and then additional 48 epochs. During one epoch the entire training dataset is processed by the network. After each epoch the parameters are updated and the network learns how to make better predictions. However, the using of too many epochs should be avoided as it can cause overfitting, which means that the network learns the noise too, and it's not generalising well to new data. A 3x3 sized kernel is used with a stride of 2. The dropout rate is set to 0.5, ensuring that the model won't become overly reliant on specific nodes when making predictions.

The learning process of the network is supervised, meaning that the model is presented with labelled data during the training. The program's goal is to find patterns by which it can distinguish between data with different labels (like 'good' or 'bad' for example). Validation was conducted using 20% of the dataset on randomly selected samples. Since the network was trained to make a difference between two types of images, the final activation function was chosen to be a sigmoid function, as it is well suited for binary classification problems. For the same reason the loss function is set for binary crossentropy function, which is typically paired with sigmoid activation functions. Optimization was conducted by the Adam optimizer [14]. This code was made with Keras with Tensorflow backend [15, 16], implemented in Python, while the training and testing was conducted on a processing unit provided by the Wigner Scientific Computing Laboratory.

## 4    Training dataset

In this study, there are two types of images we aim to distinguish between: images with small ice patches on them, and images that show none. To achieve this, HiRISE images were collected and organised in these two groups, which the program learned to separate from each other.

Given the size of the HiRISE images, they need to be chunked before loaded as input. This was done using the Mars Orbital Data Explorer Access software [17] which conducted the download from the Mars Orbital Data Explorer (ODE) site [18], the chunking process and deleting of the chunks with black pixels around the centre of the image, leaving the black borders and majority of damaged pictures out of consideration. The uniform chunk size is 1024 pixel by 1024 pixel, which is a necessary size for the surface features to be taken into consideration during the training process, not just the bright patches themselves. The images were also converted to grayscale before chunking to improve training time. Another important aspect was that only a small portion of the Martian surface is imaged in HiRISE RGB, so it's crucial for the program to recognise ice patches in black and white.

Out of the analysed 110 HiRISE images 34 were picked out for training after converting them to grayscale, 26 with small icy patches on them and 9 with none or with visible $CO_2$ sublimation pattern on the surface. Additional 38 images were used during the test run, 3 icy and 35 not. After the chunking process the dataset is expanded to 18646 images, which is sufficient for training a CNN. The chunks from the 29 icy images were then individually categorised into icy and not icy chunks to make sure the data is organised properly.

A bright patch is considered icy, if it is located on the poleward side of a shadowing landform, does not cast own shadow and has slightly diffuse edges. These were considered as selection criteria for the automated evaluation of image chunks. 252 chunks with bright patches that were difficult to identify (for example because of clouds were left out of the training process.

After chunking and categorising, the dataset of 72 images was expanded to **18646 image chunks**. Out of these, 20% showed small ice patches visible, while the remaining chunks showed no ice patch or $CO_2$ sublimation was visible on the surface (like in Figure 2, 1st column 3rd image). Figure 2 is a visualisation of a small portion (1024 x 1024 pixel, e.g. 307 x 307 m sized) of the dataset used for training.

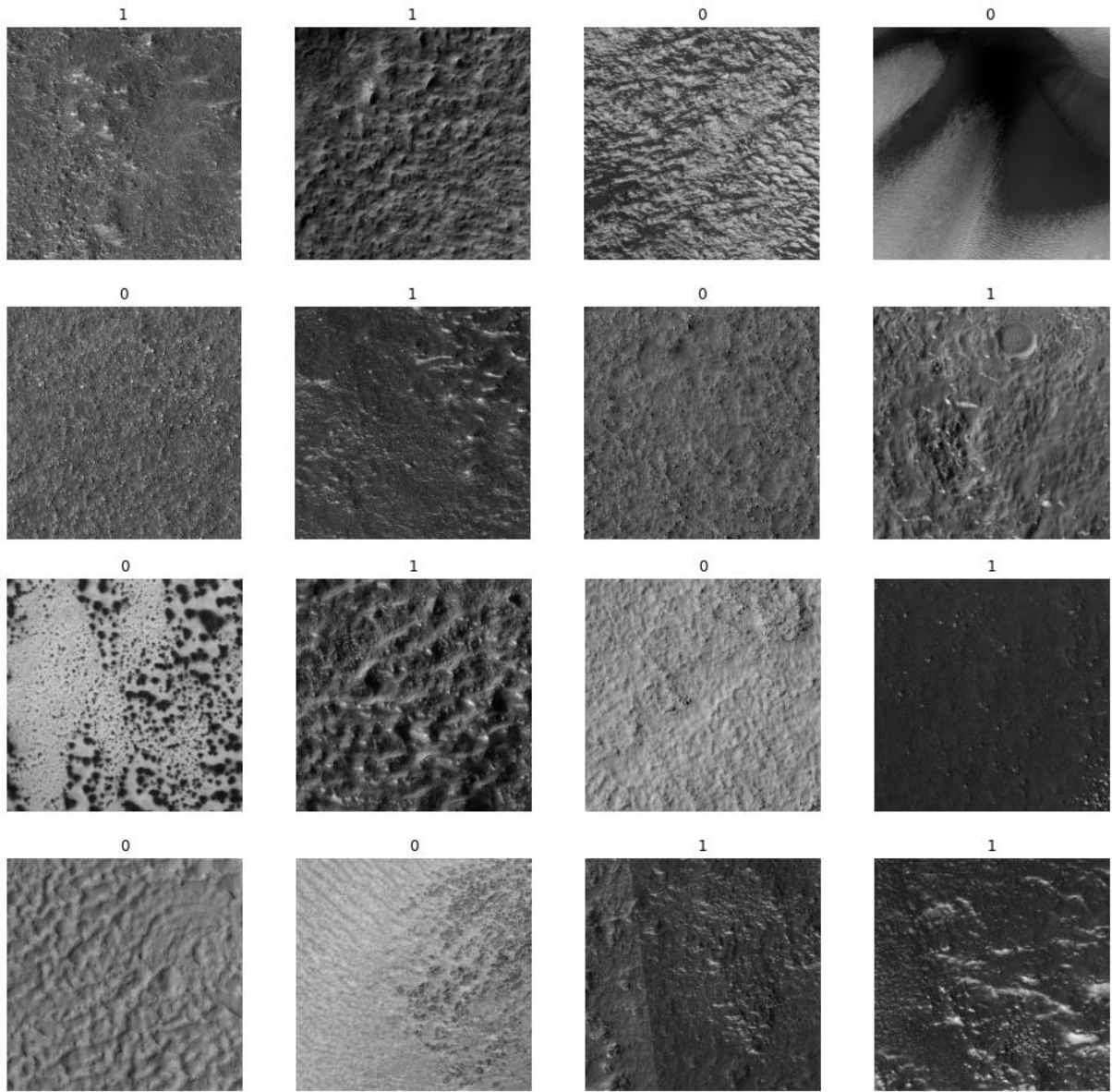

**Figure 2:** A few examples from the dataset (1024 x 1024 pixel, e.g. 307 x 307 m sized). Images marked with 1 show small ice patches, while the ones with 0 have no small ice patch visible or $CO_2$ ice is sublimating from an extended area of the surface causing dark and not bright patches.

## 5    Results

This section has the following structure: first presenting the loss and accuracy values, then presenting results of running the program on new data.

The evolution of the loss and accuracy values by the software during the epochs is shown in Figure 3. The **loss** measures the difference between the predicted output by the model and the true output (meaning how well it predicted the correct label of a given data), while **accuracy** measures the proportion of the correctly classified images from the training dataset. The gradual decrease of the loss along with the training epochs indicates that the model is improving in making better predictions over time. Occasional spikes in the curve can be caused by the randomness introduced by the dropout, when a certain percentage (in this case

50%) of randomly selected nodes were dropped out, essentially removing the connections to and from these nodes. In the later epochs the curve can be seen to decrease more gradually, which indicates that even further developments are difficult to achieve. As it can be seen, the accuracy of the predictions increases as the loss decreases, however with this size of the dataset and number of epochs, the validation loss must be also monitored to avoid the use of an over fitted model After the end of the 28th epoch, validation loss is the closest to the loss curve and starts to slowly increase after, meaning that the model is starting to over fit. The accuracy at epoch 28 is 98%, which is sufficient. Validation was conducted using 3729 random images from the training dataset.

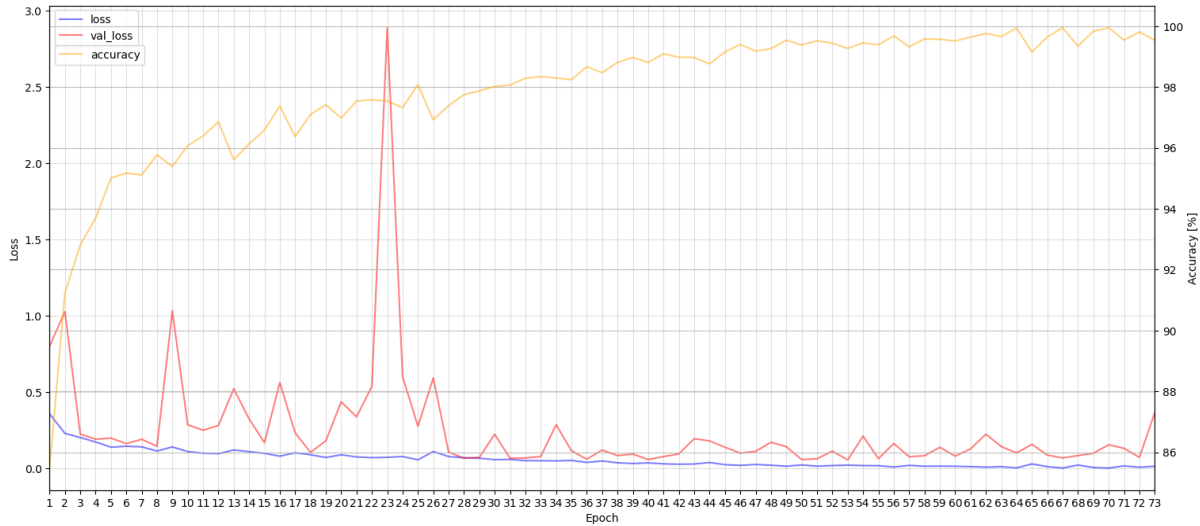

**Figure 3:** Evolution of the loss and accuracy during the training epochs. The decreasing trend of the loss indicates that the model is successfully learning to make better predictions. However, when compared to the validation loss, one can see that around epoch 30 the val_loss is starting to grow while the loss is still decreasing. This means that after epoch 30 the model is most likely starting to over fit, which is to be avoided.

The model analysed 216 new HiRISE images (cut into 13879 chunks) from the area and seasonal range of interest. Randomly selected small insets of 307 x 307 m sized areas are visualised in Figure 4 with their corresponding predictions for small ice patches being present. The program found the small icy patches with high probability. Percentages between 40-60% were considered as 'hard to identify', like the one in image 7 on Figure 4. The reason for the lower percentage could be the fact that the bright patch is located at the edge of the chunk, making it difficult for the model to recognise surface formations.

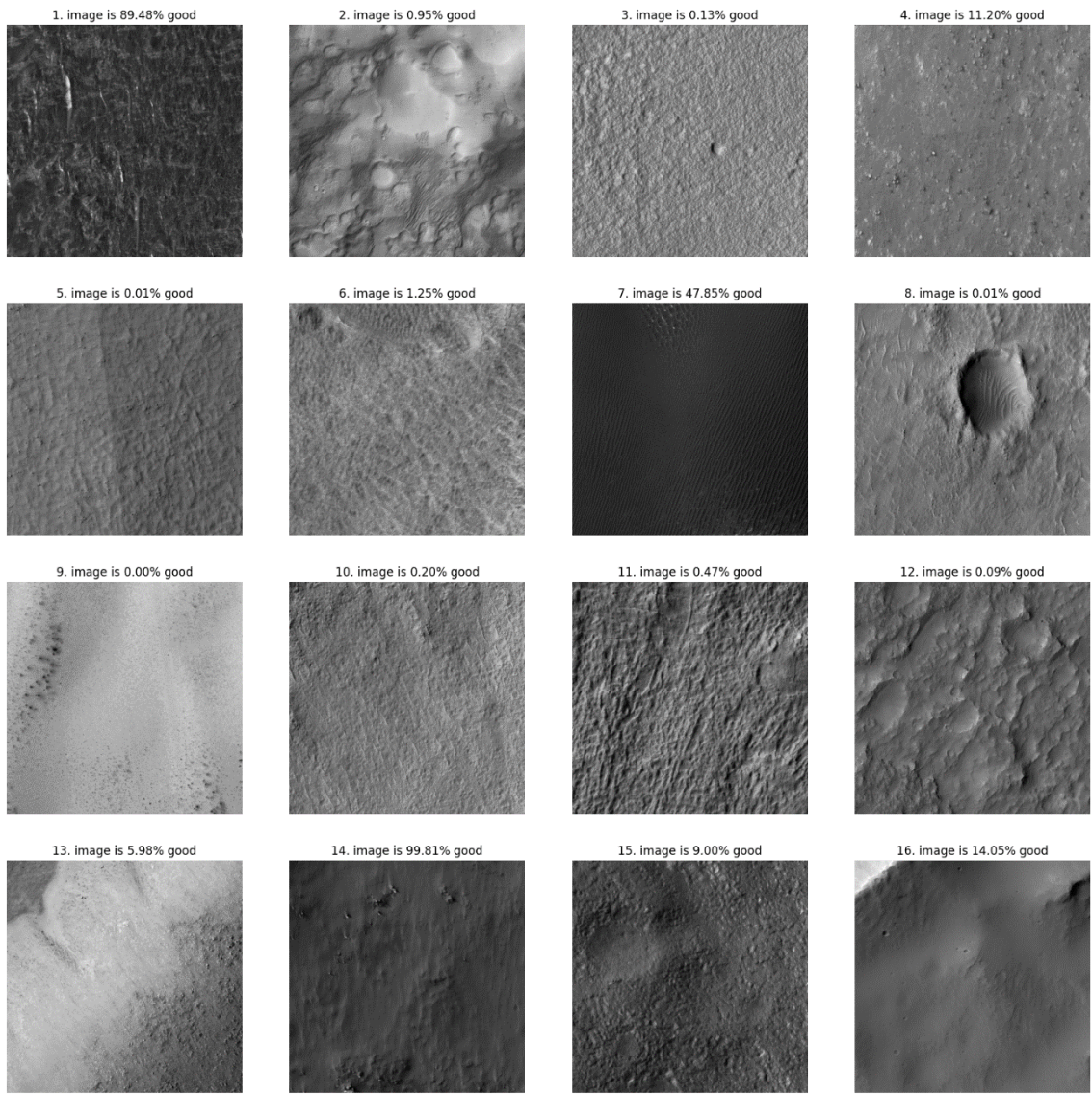

**Figure 4:** Example images and their corresponding predicted probability for having small icy patches visible on them. Above each image the predicted probability of having small ice patches on the surface is displayed in a percentage. The model was successful in finding residual ice patches on image 1 and 14, these were confirmed by manual analysis as well. The model didn't confuse light patches with ice on the remaining images

A HiRISE image was considered 'good', if the average probability for ice patch was above 60% with more than half of the chunks having at least 60% prediction for small icy patch on them. If the amount of icy chunks is less than half, and image was categorised as 'good' if the average probability was above 20%. The model categorised an image as 'hard to identify' if with more than half of the chunks being predicted as icy the average probability is between 40-60%, while the ones with less than half but more than 1 chunk considered icy the minimum percentage for this category was 1%. Anything under the minimum was considered 'bad'.

# 6 Discussion

After the initial model was trained it produced a very low accuracy (13.6%). The reason for this was that images with minor pixel anomalies (absolute bright pixels in the image that can be the result of errors during image processing or data transmission etc.) were not taken into consideration during the training. During the testing the model falsely identified these white pixels as ice patches in 5 HiRISE images and in more than 150 chunks. Further errors were commonly made by the model when a bright patch was located in the edge of the chunk, essentially cutting it in half in most cases. Bright patches located not only in shaded but in sunlit areas caused unreliable predictions too, as well as the low contrast between the surface and the brighter patch.

After retraining the model with more data and implementing the experiences of the previous attempt, the program worked well with predicting the presence or absence in commonly occurring surface forms like stone fields, smaller craters or individual smaller rocks (like in image 14). These are the most common shading forms where ice remains after the seasonal recession of the polar ice cap, therefore most of the training dataset consisted of images with such shadowing features in them. The model recognised ice correctly in 94% of the cases, but since almost half of them covered the area in too large area, the model was correct in 58% of the cases. One image predicted to be bad was qualified as good after manual analysis, however the model found the icy chunks.

# 7 Summary

Out of the 216 HiRISE images, 98 were predicted to have no chunks with small icy patches at all. The prediction was manually checked for all HiRISE images categorized as 'good' and 'hard to identify', and for around 200 chunks out of the 13879 ones from the testing dataset.

After the manual check, the classification of all of the 200 chunks was correct. This level of accuracy must be due to a specific classification criteria, which is that predictions between 40-60% are considered 'hard to identify'. After the manual check, 29 out of the 31 images classified as 'good' by the model showed ice, but only 18 of these images were actually sufficient. The rest of these icy images showed areas where $CO_2$ is sublimating, meaning that most likely the whole area in the image is still covered with $CO_2$ ice as well. For the two other images the colour information was needed to manually classify them as 'bad'.

Out of the 39 images considered 'hard to identify', 11 was good after the manual check. One of the images classified 'bad' turned out to be good after the manual check, The model found icy chunks with high probability, however the average prediction was only 0.03%.

The model presented high accuracy predictions in recognising HiRISE images with ice patches present. Therefore using the program in the search for residual water ice patches may be beneficial, as it significantly decreases the number of HiRISE images that need to be analysed manually.

## Acknowledgement

The authors thank the Wigner Scientific Computing Laboratory for their support.

Many thanks to Bíró Gábor as well for the insightful discussions on machine learning and convolutional neural networks.